\documentclass{article}
\usepackage{color}
\usepackage{graphicx}
\usepackage{amsmath, amsthm}
\usepackage{mathtools}
\usepackage{natbib}
\usepackage{url}
\usepackage{soulutf8}
\RequirePackage[colorlinks,citecolor=blue,urlcolor=blue]{hyperref}
\usepackage{xcolor}
\usepackage{chngcntr}
\usepackage{subfig}
\usepackage{hyperref}
\usepackage{algorithm}
\usepackage[noend]{algpseudocode}
\usepackage{setspace}
\newtheorem{theorem}{Theorem}
\newtheorem{corollary}{Corollary}
\usepackage{xfrac}
\definecolor{forest}{rgb}{0.133,0.545,0.133}
\usepackage{multirow}
\usepackage{amsfonts}
\usepackage{enumitem}

\evensidemargin=\oddsidemargin
\usepackage{etoolbox}

\newif\ifabbreviation
\pretocmd{\thebibliography}{\abbreviationfalse}{}{}
\AtBeginDocument{\abbreviationtrue}

\begin{document}
	\newcommand{\bb}{\boldsymbol{\beta}}

	\title{An Economical Approach to Design with Precision Criteria}


	\author{Luke Hagar\footnote{Luke Hagar is the corresponding author and may be contacted at \url{luke.hagar@ubc.ca}.} \hspace{35pt} Nathaniel T. Stevens$^{\dagger}$ \bigskip \\ 
 $^*$\textit{Department of Statistics, The University of British Columbia} \\ $^{\dagger}$\textit{Department of Statistics \& Actuarial Science, University of Waterloo}}

	\date{}

	\maketitle

	\begin{abstract}

Estimation frameworks for statistical inference are preferred to hypothesis testing when quantifying uncertainty and precise estimation are more valuable than binary decisions about statistical significance. Study design for estimation-based investigations often uses precision criteria to select sample sizes that control the length of interval estimates with respect to a sampling distribution. In this paper, we formally define the length probability distribution which characterizes the probability of obtaining a sufficiently narrow interval estimate as a function of the sample size. This distribution can then be used to determine the smallest sample size needed to ensure an interval estimate is sufficiently narrow. We prove that this distribution is approximately normal in large-sample settings for many data generation processes. However, this approximate normality may not hold for studies with moderate sample sizes, particularly when incorporating prior information or obtaining asymmetric interval estimates. Thus, we also propose an efficient simulation-based approach that estimates the sampling distribution of interval estimate lengths at only two sample sizes. Our methodology provides a unified framework for design with precision criteria in Bayesian and frequentist settings with parametric, semiparametric, and nonparametric inference. We illustrate the broad applicability of this framework with various examples.
\end{abstract}

		\bigskip

		\noindent \textbf{Keywords:}
		Credible intervals; confidence intervals; sample size determination; study design; the Bernstein-von Mises theorem

	\maketitle

	\baselineskip=19.5pt


	\section{Introduction}\label{sec:intro}

    Practitioners across disciplines design studies to collect data that meaningfully inform decision making in the presence of uncertainty. Different study objectives -- such as assessing the plausibility of hypotheses or precisely estimating quantities of interest -- motivate distinct approaches to study design. While many studies are designed with the goal of rejecting a null hypothesis, the null hypothesis significance testing framework has been scrutinized for its association with the scientific replicability crisis \citep{open2015estimating}. The American Statistical Association replied to these concerns by releasing an official statement on $p$-values \citep{wasserstein2016asa} that challenged the statistical community to develop alternatives and extensions to the traditional hypothesis testing framework. This statement prompted various responses including a special issue of \textit{The American Statistician}  that discussed ``moving to a world beyond $p < 0.05$'' \citep{wasserstein2019moving}. 
    
    Amid the outpouring of work that followed, several contributions proposed methods for data analysis that leverage interval estimates for some scalar target of inference $\theta$ to inform decision making in Bayesian or frequentist settings \citep{blume2019introduction, goodman2019proposed, matthews2019moving}. In both statistical paradigms, sample size determination within the estimation framework is used to ensure these interval estimates are precise enough to reliably inform decision making. In standard frequentist settings, the required sample size $n$ can be found using explicit formulas or by numerically solving equations \citep{mchugh1961confidence, thompson1987sample, dattalo2008determining, riley2021minimum, mondal2024review}. Analytical formulas for the sample size may also be available in simple Bayesian settings with conjugate priors \citep{adcock1987bayesian,joseph1997bayesian}. However, required sample sizes must often be found using simulation in complex Bayesian and frequentist contexts \citep{wang2002simulation, santis2004two}. 
    
   Despite the differences between statistical paradigms, the asymptotic equivalence between Bayesian credible sets of credibility level $1 - \alpha$ and frequentist confidence sets of confidence level $1 - \alpha$ holds in many settings \citep{vaart1998bvm, miller2021asymptotic}. However, sample size determination in the estimation framework lacks a unified approach that spans both Bayesian and frequentist settings. Regardless of the paradigm used for data analysis, the data are unknown prior to collection. To assess the precision of the interval estimate for $\theta$ at the design stage, a data generation process is specified.  In Bayesian contexts, \citet{gubbiotti2011bayesian} defined two methodologies for choosing this process: the conditional and predictive approaches. The predictive approach -- but not the conditional one -- accounts for uncertainty in the data generation process. The conditional approach is typically used in frequentist sample size calculations, and the predictive approach is commonly viewed as more consistent with the Bayesian paradigm. However, the predictive approach is also used in frequentist contexts \citep{chuang2006sample, ren2014assurance}, and Bayesian studies can be designed under the conditional approach \citep{fda2019adaptive,stevens2022cpm, deng2024metric}. There is thus a material need for design methodology that accommodates the conditional and predictive approaches in both statistical paradigms.   

To accommodate Bayesian and frequentist approaches, we broadly introduce precision criteria for interval estimates. We henceforth use the term CI for statements that apply to \emph{both} confidence intervals and credible intervals. Our methods accommodate various types of CIs in parametric, semiparametric, and nonparametric settings.  For now, we denote the length of the $100 \times (1-\alpha)\%$ CI for the target of inference $\theta$ as $L_{1-\alpha}$. We later emphasize that this CI is a function of the data. Precision criteria generally select a sample size $n$ to ensure a large probability that $L_{1-\alpha}$ is sufficiently narrow with respect to the sampling distribution for the future data. The length probability criterion (LPC) was defined to formalize this objective in Bayesian settings \citep{santis2004two, brutti2014bayesian}. Formally, for some target length $l$, the length criterion is simply $L_{1-\alpha} \le l$, and the LPC is $\Pr(L_{1-\alpha} \le l) \ge q$ for some  probability $q \in (0,1)$. In a Bayesian \emph{or} frequentist setting, the LPC could be applied to select the smallest sample size $n$ such that $\Pr(L_{1-\alpha} \le l) \ge q$. This focus on controlling the length of CIs is not to the exclusion of the frequentist notion of coverage and its Bayesian analogs \citep{gustafson2012behaviour}, which should approach the nominal level of $1 - \alpha$ as $n \rightarrow \infty$ under standard regularity conditions \citep{bickel1981some,lo1987large, vaart1998bvm, miller2021asymptotic}. 

In this work, we define a distribution for the LPC with respect to the sample size $n$ and explore its theoretical properties. We call this distribution the \emph{length probability distribution} (abbreviated as the LP distribution). The LP distribution is formally defined so that its $q$-quantile is the smallest sample size $n$ such that $L_{1-\alpha}$ will have length at most $l$ with probability at least $q$ (i.e., such that $\Pr(L_{1-\alpha} \le l) \ge q$). This construct broadly accommodates Bayesian and frequentist CIs under both the conditional and predictive approaches to specify sampling distributions. 

By estimating the LP distribution and its quantiles, practitioners can select sample sizes for scientific studies that attain precision criteria. To support flexible study design based on precision criteria, sample sizes that satisfy the LPC can be found using intensive simulation \citep{wang2002simulation}. In general, many samples of size $n$ are simulated to estimate the sampling distribution of CI lengths, and the proportion of samples for which the CI length is at most $l$ estimates the probability the length criterion is satisfied for that sample size. That process is repeated for each $n$ considered until the desired quantile of the LP distribution is found. Although simulation-based procedures can be generally applied to design studies with various statistical models and CI types, such naive procedures are computationally prohibitive and wastefully ignore information provided by the large-sample theory that pertains to interval estimates (see e.g., \citet{vaart1998bvm}). It would therefore be ideal to develop a framework for design with precision criteria that leverages CI theory to efficiently guide flexible simulation-based procedures. Such an economical framework to determine the minimum sample size $n$ that satisfies the LPC would meaningfully expedite design with precision criteria in Bayesian and frequentist contexts. 

In this paper, we propose two methods to estimate the LP distribution across statistical paradigms. The first method analytically approximates the LP distribution using asymptotic theory. The second method leverages this asymptotic theory to inform an efficient method to estimate the LP distribution based on simulation results conducted at only two sample sizes; this method is generally applicable and flexibly accommodates moderate sample size settings where asymptotic approximations may be unsuitable.  While our design framework proposed here is theoretically intricate, its implementation is straightforward, promoting an economical and broadly useful approach to simulation-based sample size determination with precision criteria.

The remainder of this article is structured as follows. We introduce background information and notation in Section \ref{sec:prelim}. In Section \ref{sec:ssd}, we state the conditions under which the LP distribution is approximately normal and prove this result as a theorem.  In Section \ref{sec:proxy}, we develop a theoretical proxy to the sampling distribution of CI lengths, and we propose a method in Section \ref{sec:alg} that adapts this theory to determine the minimum sample size $n$ that satisfies the LPC. Our unified approaches to estimate the LP distribution can be utilized in Bayesian and frequentist settings under the conditional or predictive approach. In Section \ref{sec:study}, we conduct numerical studies for several examples to assess the accuracy with which our proposed methods estimate the LP distribution and recommend sample sizes. Section \ref{sec:conclusion} concludes with a summary and discussion of extensions to this work. Additional examples provided in an online supplement further illustrate the broad applicability of our methodology. 

  \section{Preliminaries}\label{sec:prelim}


    Our design framework denotes data from a random, future sample as $\boldsymbol{W}^{_{(n)}}$, where the sample size $n$ is the number of independent sampling units (ISUs). The observed data are denoted by $\boldsymbol{w}^{_{(n)}}$. For non-clustered data, each independent observation is an ISU and in settings with clustered data, multiple dependent observations may comprise a single ISU. This framework broadly accommodates studies based on scalar observations and settings with additional covariates, including those that may encode information about assignment to ISUs or indicate the presence of censoring. 
    
    We assume that each ISU in $\boldsymbol{W}^{_{(n)}}$ is generated independently according to a data generation process $G$. The target of inference $\theta$ is specified as a functional $T(\cdot)$ of this process: $\theta = T(G)$. In an estimation framework, we aim to draw inference using a two-sided $100 \times (1-\alpha)\%$ CI for $\theta$. We henceforth denote the length of the $100 \times (1-\alpha)\%$ CI as $L_{1-\alpha}(\boldsymbol{W}^{_{(n)}})$ for a future sample of $n$ ISUs; the length of the corresponding CI for an observed sample is $L_{1-\alpha}(\boldsymbol{w}^{_{(n)}})$. This notation now emphasizes that the CI length and its sampling distribution are functions of the data and the confidence or credibility level $1 - \alpha$. 

Algorithm \ref{alg.int} details a simulation-based procedure to estimate the sampling distribution of CI lengths at a given sample size $n$. We use $m$ simulation repetitions to estimate the sampling distribution. Flexible semiparametric and nonparametric data generation processes may be considered for $\boldsymbol{W}^{_{(n)}}$, but this process $G_r$ must be fully specified to simulate data in simulation repetition $r = 1, \dots, m$. In addition to generating response variables, $G_r$ may need to specify how additional covariates or ISU sizes are generated. For Bayesian analyses only, a prior distribution $p(\boldsymbol{\theta})$ for the target of inference must also be chosen. 

The final input that we specify for Algorithm \ref{alg.int} is $\mathcal{G}$, a data generation family that may incorporate uncertainty in the data generation process; various examples are provided in Section \ref{sec:study}.  If uncertainty is incorporated into $\mathcal{G}$, it is common to fix the data generation model but allow (a subset of) its parameters to vary across simulation repetitions according to some probability distribution. The data generation family $\mathcal{G}$ is respectively degenerate and nondegenerate under the conditional and predictive approaches \citep{gubbiotti2011bayesian}. In particular, all data generation processes  $\{G_r\}_{r=1}^m$  in Line 3 of Algorithm \ref{alg.int} are identical under the conditional approach. Line 3 also generates a sample $\boldsymbol{w}^{_{(n)}}_{r}$ in each simulation repetition given the data generation process $G_r$ drawn from $\mathcal{G}$.

\begin{algorithm}
\caption{Sampling Distribution Estimation}
\label{alg.int}

\begin{algorithmic}[1]
\setstretch{1}
\Procedure{Estimate}{$T(\cdot)$, $\alpha$, $p(\theta)$, $n$, $m$, $\mathcal{G}$}
\For{$r$ in 1:$m$}
    \State Generate $G_{r} \sim \mathcal{G}$ and $\boldsymbol{w}^{_{(n)}}_{r} \sim G_r$ 
    \State Compute estimate $\widehat{L}_{1-\alpha}(\boldsymbol{w}^{_{(n)}}_{r})$ 
    \EndFor
    \State \Return $\{\widehat{L}_{1-\alpha}(\boldsymbol{w}^{_{(n)}}_{r})\}_{r = 1}^m$
\EndProcedure

\end{algorithmic}
\end{algorithm}

 The collection of estimates $\{\widehat{L}_{1-\alpha}(\boldsymbol{w}^{_{(n)}}_{r})\}_{r = 1}^m$ in Line 5 of Algorithm \ref{alg.int} is used to estimate the sampling distribution of CI lengths under $\mathcal{G}$. The relevant $100 \times (1-\alpha)\%$ CIs for $\theta$ can be estimated using a multitude of Bayesian or frequentist methods based on parametric, semiparametric, or nonparametric inference; more details are provided in Section \ref{sec:ssd}. The probability of satisfying the length criterion is estimated as
\begin{equation}\label{eq:oc.est}
\dfrac{1}{m}\sum_{r=1}^m\mathbb{I}\left\{\widehat{L}_{1-\alpha}(\boldsymbol{w}^{_{(n)}}_{r}) \le l \right\}.
\end{equation} 
The probability of satisfying the length criterion, defined given fixed $l$ and $\alpha$, approaches 1 as the sample size $n \rightarrow \infty$ under standard regularity conditions (see e.g, \citet{bickel1981some,lo1987large, vaart1998bvm, miller2021asymptotic}). The optimal sample size $n$ is selected as the smallest integer such that the estimate in (\ref{eq:oc.est}) is at least $q$.

The optimal choice for $n$ is informed by exploring the sampling distribution of CI lengths to estimate the probability of satisfying the length criterion in (\ref{eq:oc.est}). However, the procedure in Algorithm \ref{alg.int} requires independent implementation for each $n$ value we consider. This process can be computationally intensive, but we could reduce the computational burden by repurposing the results from Algorithm \ref{alg.int} for previously considered sample sizes to estimate (\ref{eq:oc.est}) for new $n$ values. This process would allow us to explore new sample sizes without conducting additional simulations. We propose such a method that combines simulation and large-sample theory to design studies with precision criteria in this paper and begin its development in Section \ref{sec:proxy}. But first, we formally define the LP distribution and prove that it is approximately normal under certain conditions in Section \ref{sec:ssd}, which allows us to contrast the performance of our economical simulation-based method with results based solely on large-sample theory.
		

	\section{Approximate Normality of the Length Probability Distribution}\label{sec:ssd}

       The LP \emph{distribution} is defined such that when the sample size $n$ is taken as its $q$-quantile, the resulting $100\times(1-\alpha)\%$ CI will have length at most $l$ with probability at least $q$. We let the LP \emph{curve} be the cumulative distribution function (CDF) of the LP distribution. More precisely, let $F_l(n)$ be the LP \emph{curve} for fixed $l$ and $1 - \alpha$ such that
\begin{equation}\label{eq:lp.curve}
    F^{-1}_l(q) := \text{inf}\{n \in \mathbb{Z}^{+}: \Pr(L_{1-\alpha}(\boldsymbol{W}^{_{(n)}}) \le l) \ge q \}.
\end{equation}
That is, $F_l(n) = \Pr(L_{1-\alpha}(\boldsymbol{W}^{_{(n)}}) \le l)$ for all $n \in \mathbb{Z}^+$. This definition of the LP curve assumes the ISUs in $\boldsymbol{w}^{_{(n)}}_{r}$ are generated according to the process $G_r$ such that $G_{r} \sim \mathcal{G}$. We use the term LP curve because the plot of $\Pr(L_{1-\alpha}(\boldsymbol{W}^{_{(n)}}) \le l)$ as a function of the sample size $n$ is an analog to the power curve used in null hypothesis significance testing.

In this section, we provide theoretical justification for the approximate normality of the LP distribution under the conditional approach where the data generation family $\mathcal{G}$ is degenerate. This normality is useful because it facilitates fast analytical sample size calculations that we improve upon with the broader methodology described in Sections \ref{sec:proxy} and \ref{sec:alg}. We later describe how this approximate normality can also be exploited under the predictive approach.

To ensure the approximate normality of the LP distribution, we require that two broadly applicable asymptotic results hold true. The \emph{first} necessary asymptotic result concerns the sampling distribution of the estimator $\hat{\theta}^{_{(n)}}$ in frequentist settings and the posterior distribution of $\theta$ in Bayesian ones. For large $n$, we require that this sampling or posterior distribution is approximately
      \begin{equation}\label{eq:norm1}
  \mathcal{N}\left(\theta_r, \Lambda^2_r/n\right),
\end{equation}
where $\theta_r$ is the value that $\hat{\theta}^{_{(n)}}$ or the posterior distribution of $\theta$ converges to in probability under $G_r$ as $n \rightarrow \infty$, and $\Lambda^2_r$ is a limiting unscaled variance. We emphasize that $\Lambda_r$ is a constant under the conditional approach since $\mathcal{G}$ is degenerate. When the data generation process $G_r$ is known, $\Lambda_r$ can be computed analytically or numerically.  In practice, $\Lambda_r$ must instead be estimated using an estimator $\hat{\Lambda}_r^{_{(n)}}$ that is informed by an observed sample $\boldsymbol{w}^{_{(n)}}$. For fixed $\alpha$, it follows from (\ref{eq:norm1}) that
    \begin{equation}\label{eqn:length}
    L_{1 - \alpha}(\boldsymbol{W}^{_{(n)}}) \approx \dfrac{2 z_{1-\alpha/2}}{\sqrt{n}}\times \hat{\Lambda}_r^{_{(n)}},
    \end{equation} 
    where $z_{1-\alpha/2}$ is the upper $\alpha/2$ quantile of $\mathcal{N}(0,1)$. 
    
    The \emph{second} necessary asymptotic result pertains to the sampling distribution of $\hat{\Lambda}_r^{_{(n)}}$. For large $n$, we require that this sampling distribution is approximately
      \begin{equation}\label{eq:norm2}
  \mathcal{N}\left(\Lambda_r, \tau^2_r/n\right),
\end{equation}
where $\tau^2_r$ is the limiting unscaled variance for this distribution according to $G_r$. 

The results in (\ref{eq:norm1}) and (\ref{eq:norm2}) hold true in many settings. While not an exhaustive list, those results generally hold true when one of the following sets of regularity conditions are satisfied: those for the asymptotic normality of the maximum likelihood estimator \citep{vaart1998bvm} or M-estimator \citep{huber2009robust}, those for the martingale central limit theorem \citep{fleming2005counting}, those for the Bernstein-von Mises theorem \citep{vaart1998bvm} or its semiparametric analogs \citep{miller2021asymptotic}, or those for the bootstrap in frequentist  \citep{bickel1981some} or Bayesian \citep{lo1987large} settings for various targets of inference. There are, however, rare situations where the relevant regularity conditions and hence the results in (\ref{eq:norm1}) and (\ref{eq:norm2}) do not hold true. For instance, the sampling distribution of the maximum likelihood estimator for $\theta$ in a $\mathcal{U}[0, \theta]$ distribution is asymptotically exponential. We verify that these results hold true for our varied examples in Section \ref{sec:study}. If none of the regularity conditions listed above can be applied for a given study, we encourage practitioners to confirm whether the results hold true by consulting the relevant theory or conducting a small-scale numerical study. 
    
    The results in (\ref{eqn:length}) and (\ref{eq:norm2}) imply that for large sample sizes $n$, \begin{equation}\label{eqn:impa}
    L_{1 - \alpha}(\boldsymbol{W}^{_{(n)}}) ~\dot\sim~ \mathcal{N}\left(\dfrac{2 z_{1-\alpha/2}\Lambda_r}{\sqrt{n}}, \dfrac{4 z^2_{1-\alpha/2} \tau_r^2}{n^2}\right).
    \end{equation}
    Theorem \ref{thm1} uses the approximate sampling distribution for $L_{1 - \alpha}(\boldsymbol{W}^{_{(n)}})$ in (\ref{eqn:impa}) to prove that the LP distribution is approximately normal under the conditional approach. This approximate normality holds when the interval length $l$ decreases to 0. As $l \rightarrow 0^+$, the sample sizes $n$ that satisfy the LPC increase to $\infty$. 
    
        \begin{theorem}\label{thm1}
For $G_r$ from a degenerate data generation family $\mathcal{G}$, let the results in (\ref{eq:norm1}) and (\ref{eq:norm2}) hold true such that $ 0 < \Lambda_r^2 < \infty$ and $0 < \tau_r^2 < \infty$. Let $F_l(n)$ be the LP curve for $L_{1-\alpha}(\boldsymbol{W}^{_{(n)}})$ with fixed $l$ and $1 - \alpha$. Let $\mu_l =  4 z^2_{1-\alpha/2}\Lambda_r^2/l^2$, $\sigma_l = 4z_{1-\alpha/2} \tau_r/l$, and $\Phi(\cdot)$ be the standard normal CDF. Then, 
	\begin{equation*}\label{eqn:thm}
	\lim_{l\rightarrow 0^+}	\sup_{n \in \mathbb{Z}^+} \left\lvert F_l(n) - \Phi\left(\dfrac{n - \mu_l}{\sigma_l}\right) \right\rvert \xrightarrow{P} 0.
	\end{equation*} 
	
\end{theorem}

We prove Theorem \ref{thm1} in Appendix A of the supplement. This theorem imposes constraints on $\Lambda_r$ and $\tau_r$ to ensure the normal distribution that approximates the LP distribution for a given length $l$ has finite mean $\mu_l > 0$ and standard deviation $\sigma_l > 0$. If it is too cumbersome to obtain $\Lambda_r$ and $\tau_r$ analytically, their values can be estimated using a small-scale simulation as follows. First, simulate $\boldsymbol{w}^{_{(n)}}$ with large $n$ to estimate $\Lambda_r$ as $L_{1 - \alpha}(\boldsymbol{w}^{_{(n)}}) \div 2z_{1-\alpha/2}/\sqrt{n}$. Given this estimate $\tilde{\Lambda}_r$, select an initial sample size of $n = \lceil4z^2_{1 - \alpha/2}\hat{\Lambda}_r^2/l^2\rceil$. At this sample size, run a small simulation (e.g., obtain 100 CI lengths). Lastly, divide the standard deviation of $\{L_{1 - \alpha}(\boldsymbol{w}_s^{_{(n)}})\}_{s=1}^{100}$ by $2z_{1-\alpha/2}/n$ to get $\tilde{\tau}_r$, and update $\tilde{\lambda}_r$ as the average of $\{L_{1 - \alpha}(\boldsymbol{w}_s^{_{(n)}}) \div 2z_{1-\alpha/2}/\sqrt{n}\}_{s=1}^{100}$. The estimates $\tilde{\lambda}_r$ and $\tilde{\tau}_r$ can be substituted into Theorem \ref{thm1} to get $\mu_l$ and $\sigma_l$. For large $n$ corresponding to small target lengths $l$, the values for $\mu_{l}$ and $\sigma_{l}$ from Theorem \ref{thm1} can be used to determine quantiles of the LP distribution under the conditional approach as stated in Corollary \ref{cor1}.

\begin{corollary}\label{cor1}
    Let $\mu_l$ and $\sigma_l$ be defined as in Theorem \ref{thm1}. If we obtain a sample of size $n = \mu_l + z_{q}\sigma_l$ according to a given data generation process $G_r$, the probability that the $100\times(1-\alpha)\%$ CI for $\theta$ satisfies the length criterion approaches $q$ as $l \rightarrow 0^{+}$.
\end{corollary}

Corollary \ref{cor1} can be extended for use with the predictive approach by approximating the LP distribution as a mixture of normal distributions. In particular, one can obtain $m$ data generation processes $G_r \sim \mathcal{G}$. Theorem \ref{thm1} can be applied to obtain $\mu_{l, r}$ and $\sigma_{l, r}$ for each repetition $r$. When $m$ is sufficiently large, the LP distribution can then be estimated as the mixture of the $\{\mathcal{N}(\mu_{l, r}, \sigma^2_{l, r}) \}_{r=1}^m$ distributions. Note that similarly to how asymptotic normality is exploited for finite samples, we do not require that $l = 0$ to invoke Theorem \ref{thm1}. However, it is not trivial to determine how small $l$ must be to yield a sample size $n$ that is large enough to apply the asymptotic result in Theorem \ref{thm1} for a given study. Corollary \ref{cor1} is therefore most useful in the limiting case with large sample sizes. Nevertheless, the theoretical results from this section are valuable. In Section \ref{sec:proxy}, we leverage insights from this section to develop a method that prompts better sample size recommendations in moderate sample size scenarios -- particularly in settings with asymmetric CIs and Bayesian contexts with informative priors.


    \section{A Proxy to the Sampling Distribution of CI Lengths}\label{sec:proxy}

     Flexible, simulation-based design methods that control the length of CIs require that we estimate the sampling distribution of CI lengths for various sample sizes $n$. We approximate this sampling distribution by generating data $\boldsymbol{w}^{_{(n)}}$ using the straightforward process in Algorithm \ref{alg.int}. For theoretical development, we create a proxy for the sampling distribution of the logarithm of the CI lengths. These proxies substantiate the theory that underpins our proposed methodology, but they are not used in our design methods proposed in Section \ref{sec:alg}. 
     
     The delta method and the result in (\ref{eqn:impa}) give rise to an approximate sampling distribution for $\kappa(\boldsymbol{W}^{_{(n)}}) = \log (L_{1 - \alpha}(\boldsymbol{W}^{_{(n)}}))$. For simplicity, the notation $\kappa(\boldsymbol{W}^{_{(n)}})$ does not incorporate the confidence or credibility level $1 - \alpha$. The approximate sampling distribution for $\kappa(\boldsymbol{W}^{_{(n)}})$ is
     \begin{equation}\label{eqn:logL}
    \kappa(\boldsymbol{W}^{_{(n)}}) ~\dot\sim~ \mathcal{N}\left(\log(2 z_{1-\alpha/2}{\Lambda_r}) - \dfrac{1}{2}{\log(n)}, \dfrac{\tau_r^2}{n\Lambda_r^2} \right).
    \end{equation}
    For theoretical purposes, we could simulate from this approximate sampling distribution of $\kappa(\boldsymbol{W}^{_{(n)}})~|~G = G_r$ using CDF inversion with a point $u_{r} \in [0,1]$:  
   \begin{equation}\label{eq:cdf.inv}
\kappa^{_{(n)}}_{r} = \log(2 z_{1-\alpha/2}{\Lambda_r}) - \dfrac{1}{2}{\log(n)} + \dfrac{z_{u_{r}}\tau_r\Lambda_r^{-1}}{\sqrt{n}}.
\end{equation} 
We emphasize that $\kappa^{_{(n)}}_{r}$ in (\ref{eq:cdf.inv}) depends on the point $u_{r}$ and data generation process $G_r$ through the subscript $r$. The $\{\kappa^{_{(n)}}_{r}\}_{r=1}^m$ values corresponding to points $\{u_{r}\}_{r = 1}^m \sim \mathcal{U}[0,1]$ and data generation processes $\{G_r\}_{r=1}^m \sim \mathcal{G}$ define our large-sample proxy to the sampling distribution of the logarithm of CI lengths. Under the predictive approach, there are two sources of randomness in the proxy sampling distribution of $\kappa^{_{(n)}}$. The first source accounts for uncertainty in the data generation family $\mathcal{G}$ at the design stage. The second source is related to the point $u_{r}$ used to generate from the limiting distribution in (\ref{eqn:logL}).

For theoretical purposes, we consider how this proxy sampling distribution changes as a function of the sample size. We first consider the median $\kappa^{_{(n)}}_{*}$ of the proxy sampling distribution conditional on the data generation process $G_r$. Based on (\ref{eq:cdf.inv}), this median is a linear function of $\log(n)$:
   \begin{equation}\label{eq:med}
\kappa^{_{(n)}}_{*} = \log(2 z_{1-\alpha/2}{\Lambda_r}) - \dfrac{1}{2}{\log(n)}.
\end{equation} 
We note that $\kappa^{_{(n)}}_{*}$ in (\ref{eq:med}) is not a stochastic quantity given a fixed $G_r$. We also define the logarithm of the absolute difference between each $\kappa^{_{(n)}}_{r}$ value and $\kappa^{_{(n)}}_{*}$ as
   \begin{equation}\label{eq:diff}
\log(\lvert\kappa^{_{(n)}}_{r} - \kappa^{_{(n)}}_{*}\rvert) = \log(\lvert z_{u_{r}}\rvert \tau\Lambda_r^{-1}) - \dfrac{1}{2}{\log(n)}.
\end{equation} 
 When conditioning on a particular value of $u_{r}$ given a fixed $G_r$, the value of $\log(\lvert\kappa^{_{(n)}}_{r} - \kappa^{_{(n)}}_{*}\rvert)$ in (\ref{eq:diff}) is not random. Given $u_{r}$ and $G_r$, the quantities in (\ref{eq:med}) and (\ref{eq:diff}) are therefore deterministic functions of $n$. 

Corollary \ref{cor2} follows directly from the results in (\ref{eq:med}) and (\ref{eq:diff}). It provides guidance concerning how to economically assess whether the length criterion is satisfied at a broad range of sample sizes. In particular, Corollary \ref{cor2} formally establishes that our proxy sampling distribution can be decomposed into various deterministic, linear functions of $\log(n)$; hence, exploration of the $n$-space can be achieved by estimating the sampling distribution of CI lengths at \emph{only two} values of $n$.

\begin{corollary}\label{cor2}
    Let the conditions for Theorem \ref{thm1} be satisfied.
    For a given point $u_{r} \in [0,1]$ and data generation process $G_r$, the functions $\kappa^{_{(n)}}_{*}$ in (\ref{eq:med}) and $\log(\lvert\kappa^{_{(n)}}_{r} - \kappa^{_{(n)}}_{*}\rvert)$ in (\ref{eq:diff}) are linear functions of $\log(n)$.
\end{corollary} 
We now consider the practical implications of Corollary \ref{cor2}. For the proxy sampling distributions, the linear approximations to $\kappa^{_{(n)}}_{*}$ and $\log(\lvert\kappa^{_{(n)}}_{r} - \kappa^{_{(n)}}_{*}\rvert)$ as functions of $\log(n)$ are good global approximations for large sample sizes. These linear approximations should also be locally suitable for a range of smaller sample sizes. Therefore, the quantiles of the sampling distribution of $\kappa^{_{(n)}}_{r}$ can be modeled in the $n$-space using linear functions of $\log(n)$ when $G_r$ is fixed (i.e., under the conditional approach).  We exploit and adapt these linear trends in the proxy sampling distribution to flexibly model the sampling distribution of CI lengths using linear functions of $\log(n)$ when independently simulating samples $\boldsymbol{w}^{_{(n)}}$ under the conditional or predictive approaches as in Algorithm \ref{alg.int}. This procedure allows us to \emph{empirically} estimate the intercepts and slopes in (\ref{eq:med}) and (\ref{eq:diff}). Since we only use the limiting slopes in (\ref{eq:med}) and (\ref{eq:diff}) to initialize our method, we do not require very large sample sizes to apply our methodology with the true sampling distributions in Section \ref{sec:alg}. We support this claim in Section \ref{sec:study}.
	
	\section{Economical Estimation of the Length Probability Distribution}\label{sec:alg}

    
    We now generalize the results from Theorem \ref{thm1} to develop the  method presented in Algorithm \ref{alg2} that is easily implemented and performs well with moderate to large sample sizes. Algorithm \ref{alg2} allows users to efficiently explore the sample size space to find the minimum $n$ that satisfies the LPC. We obtain this sample size recommendation by estimating the LP distribution from the LP curve. Our approach involves estimating the sampling distributions of CI lengths at only two sample sizes: $n_0$ and $n_1$. Algorithm \ref{alg2} details a general application of our methodology with the conditional approach, and we later describe potential modifications to accommodate the predictive approach. In Algorithm \ref{alg2}, we introduce the notation $\omega(a,b)$ to denote the $a^{\text{th}}$ order statistic of the collection of observations $b$.

\begin{algorithm}
\caption{Sample Size Determination with the LP Curve}
\label{alg2}

\begin{algorithmic}[1]
\setstretch{1.2}
\Procedure{LPcurve}{$T(\cdot)$, $p(\theta)$, $q$, $\alpha$, $l$, $m$, $\mathcal{G}$}
\State Let $n_0$ be $\lceil \mu_l \rceil$ corresponding to the median of $\{ \Lambda_r\}_{r=1}^m \sim \mathcal{G}$
\State Estimate $\{\widehat{L}_{1-\alpha}(\boldsymbol{w}^{_{(n_0)}}_{r})\}_{r = 1}^m$ via Algorithm \ref{alg.int} and their logarithms $\{\hat{\kappa}^{_{(n_0)}}_{r}\}_{r = 1}^m$
\State Model the median of the sampling distribution of $\kappa(\boldsymbol{W}^{_{(n)}})$ as $\hat{\kappa}_*^{_{(n)}} = \hat{\kappa}_*^{_{(n_0)}} + 0.5\log(n_0)$  $- 0.5\log(n)$,  \linebreak \hspace*{12pt} where $\hat{\kappa}_*^{_{(n_0)}}$ is the median of  $\{\hat{\kappa}^{_{(n_0)}}_{r}\}_{r = 1}^m$
\For{$r$ in $1$:$m$}
  \State Use the linear function of $\log(n)$ passing through $(\log(n_0),  \log(\lvert\hat{\kappa}^{_{(n_0)}}_{r} -\hat{\kappa}^{_{(n_0)}}_{*}\rvert))$ with slope -0.5 to  \linebreak \hspace*{27pt}  get  $\log(\lvert\hat{\kappa}^{_{(n)}}_{r} -\hat{\kappa}^{_{(n)}}_{*}\rvert)$ for  other sample sizes
 \EndFor
\State Use binary search to find $n_1$, the smallest $n$ such that $m^{-1}\sum_{r=1}^m\mathbb{I}\{\hat{\kappa}^{_{(n)}}_{r} \le \log(l)\} \ge q$
\State Estimate $\{\widehat{L}_{1-\alpha}(\boldsymbol{w}^{_{(n_1)}}_{r})\}_{r = 1}^m$ via Algorithm \ref{alg.int} and their logarithms $\{\hat{\kappa}^{_{(n_1)}}_{r}\}_{r = 1}^m$
\State Model the median of the sampling distribution of $\kappa(\boldsymbol{W}^{_{(n)}})$ such that $\kappa_*^{_{(n)}}$ is the  linear function of  \linebreak \hspace*{12pt} $\log(n)$ passing through $(\log(n_0), \hat{\kappa}_*^{_{(n_0)}})$ and $(\log(n_1), \hat{\kappa}_*^{_{(n_1)}})$
\For{$r$ in $1$:$m$}
   \State Use the linear function of $\log(n)$ passing through $(\log(n_0),  \omega(r, \{\log(\lvert\hat{\kappa}^{_{(n_0)}}_{s} -$ $\hat{\kappa}^{_{(n_0)}}_{*}\rvert)\}_{s=1}^m))$ and  \linebreak \hspace*{27pt} $(\log(n_1),  \omega(r, \{\log(\lvert\hat{\kappa}^{_{(n_1)}}_{s} -\hat{\kappa}^{_{(n_1)}}_{*}\rvert)\}_{s=1}^m))$ to get  $\log(\lvert\hat{\kappa}^{_{(n)}}_{r} -\hat{\kappa}^{_{(n)}}_{*}\rvert)$  for  other sample sizes
 \EndFor
\State Use the functions in Lines 9 to 11 to approximate the sampling distribution of  $\kappa(\boldsymbol{W}^{_{(n)}})$ and the LP \linebreak \hspace*{12pt}  curve across the $n$-space
\State Use binary search to find $n_2$, the smallest $n$ such that  $m^{-1}\sum_{r=1}^m\mathbb{I}\{\hat{\kappa}^{_{(n)}}_{r} \le \log(l)\} \ge q$
 \State \Return LP curve and $n_2$ as the recommended $n$

\EndProcedure

\end{algorithmic}
\end{algorithm}

We elaborate on several steps in Algorithm \ref{alg2} below. We choose the initial sample size $n_0$ in Line 2 as $\lceil \mu_l \rceil$ from Theorem \ref{thm1} for the median value of $\Lambda_r$ induced by $G_r \sim \mathcal{G}$. Regardless of whether the conditional or predictive approach is used, this sample size provides a suitable starting point for LP curve estimation since $n_0$ should correspond to a quantile of the LP distribution such that $q$ is not too close to 0 or 1. We emphasize that our methodology is flexible and can readily be integrated with any computational or analytical method used to compute CI lengths in Line 3. The notation $\hat{\kappa}^{_{(n)}}_{r}$ is also introduced in Line 3. These logarithms from the true sampling distribution  of $\kappa(\boldsymbol{W}^{_{(n)}})$ leverage independently generated samples $\boldsymbol{w}^{_{(n)}}_{r}$ for each simulation repetition $r$. Unlike for $\kappa^{_{(n)}}_{r}$ from the proxy sampling distribution in Section \ref{sec:proxy}, there is no relationship between the $\hat{\kappa}^{_{(n)}}_{r}$ values corresponding to two different sample sizes that happen to have the same index $r$.


To select the second sample size $n_1$, we construct a linear approximation to $\kappa_*^{_{(n)}}$, the median of the sampling distribution of $\kappa(\boldsymbol{W}^{_{(n)}})$, as a function of $\log(n)$ using the limiting slope of -0.5 from (\ref{eq:med}) in Line 4 of Algorithm \ref{alg2}. This line passes through the point $(\log(n_0), \hat{\kappa}_*^{_{(n_0)}})$. In Lines 5 and 6, we also construct linear approximations with limiting slopes of -0.5 from (\ref{eq:diff}) to estimate $\log(\lvert\kappa^{_{(n)}}_{r} -\kappa^{_{(n)}}_{*}\rvert)$ for new sample sizes. If we retain the sign of $\hat{\kappa}^{_{(n_0)}}_{r} -\hat{\kappa}^{_{(n_0)}}_{*}$ for each simulation repetition $r$, then we can use the linear approximations in Lines 4 to 6 to estimate the sampling distribution of $\kappa(\boldsymbol{W}^{_{(n)}})$ throughout the $n$-space. However, the limiting slopes of -0.5 may not be accurate for moderate $n$ since (\ref{eq:med}) and  (\ref{eq:diff}) rely on large-sample results. Thus, we only use those slopes in this initial phase of our method. The probability of satisfying the length criterion in Line 7 is quickly estimated based on the linear approximations -- not by computing new CI lengths based on data generated from additional simulation runs. 

In Lines 9 to 11 of Algorithm \ref{alg2}, we model the sampling distribution of CI lengths across the $n$-space using linear approximations that are less reliant on large-sample results. These approximations use independent estimates of the sampling distribution at the sample sizes $n_0$ and $n_1$, and they exploit the linear trends in the proxy sampling distribution discussed in Section \ref{sec:proxy}. For the true sampling distributions, we sort the logarithms of the absolute differences between $\hat{\kappa}^{_{(n)}}_{r}$ and $\hat{\kappa}^{_{(n)}}_{*}$ estimated at $n_0$ and $n_1$ and construct linear approximations using the same order statistic at both sample sizes. This approach is suitable when the true value of $\Lambda_r$ is similar for all $G_r \sim \mathcal{G}$. When $\mathcal{G}$ is nondegenerate under the predictive approach, the process in Lines 8 to 11 can be modified: we instead split the logarithms of the CI lengths for each sample size into subgroups based on the order statistics of their $\Lambda_r$ values before constructing the linear approximations. 

In Line 13 of Algorithm \ref{alg2}, we repeat the process in Line 7 with our improved linear approximations to obtain the final sample size recommendation $n_2$. These linear approximations yield unbiased estimates of the LP curve at the sample sizes $n_0$ and $n_1$. Thus, the suitability of LP curve estimation and the corresponding sample size recommendation $n_2$ only relies on the accuracy of the \emph{empirically} estimated slopes. We compare the performance of Algorithm \ref{alg2} with the purely analytical method to estimate the LP distribution from Section \ref{sec:ssd} when considering study design for several examples in Section \ref{sec:study}.
    
	\section{Illustrative Examples}\label{sec:study}

\subsection{Measurement Method Comparison}\label{sec:study.poa}

We first assess how well our methodology estimates the LP distribution in the context of measurement method comparison. This example concerns a parametric analysis in the frequentist paradigm with repeated measurements. In this subsection, we plan to compare two measurement methods that measure systolic blood pressure (in mmHg). This example is based on a real study described in \citet{bland1999measuring} that compares reference and new measurement methods. The results in (\ref{eq:norm1}) and (\ref{eq:norm2}) hold true for this example because the conditions for the asymptotic normality of the maximum likelihood estimator \citep{vaart1998bvm} are satisfied for the model described below.

We suppose that the following linear mixed-effects structural model characterizes the data and relates measurements by the two methods:
\begin{equation} \label{eq:msdata}
    \begin{split}
       W_{i1k} &= S_i + M_{i1k} \\
    W_{i2k} &= \gamma + \beta S_i + M_{i2k},
    \end{split}
\end{equation}
where $i = 1, 2, \dots, n$ indexes the participants, $j = 1, 2$ respectively correspond to the reference and new measurement method, and $k = 1, 2, \dots, K$ indexes the replicate measurements. Thus, $W_{ijk}$ is a random variable that represents the \emph{logarithm} of the future value of method $j$'s $k^{\text{th}}$ measurement of participant $i$. We use a logarithmic transformation in model (\ref{eq:msdata}) because distributions for the blood pressure data from \citet{bland1999measuring} are right skewed. $S_i$ is a random variable that represents the unknown true value of the measurand for subject $i$ on the logarithmic scale. In this example, we make a common assumption that $S_i \sim \mathcal{N}(\mu, \sigma^2_s)$ and subjects are sampled randomly from the target population. $M_{ijk}$ is a random variable that represents the measurement error of method $j = 1, 2$. We assume that the $M_{ijk}$ are independent of each other and of $S_i$, and that they follow a $\mathcal{N}(0, \sigma^2_j)$ distribution, where $\sigma_j$ quantifies the measurement variation of method $j$. 

We note that the mean and variance parameters associated with the model (\ref{eq:msdata}) are distinct from $\mu_l$ and $\sigma_l$, the parameters that characterize the normal approximation to the LP distribution from Theorem \ref{thm1}. The parameters $\gamma$ and $\beta$ respectively characterize the fixed and proportional bias of the new measurement method relative to the reference one. Ultimately, $G_r$ is specified via the model (\ref{eq:msdata}) and parameters $(\gamma, \beta, \mu, \sigma_s, \sigma_1, \sigma_2)$. Each ISU is of size $2K$ since there are $K$ observations per participant for each measurement method. While ISU sizes could differ according to some probability distribution incorporated into $G_r$, we assume the ISU sizes are the same for all participants in this example, which is reasonable for measurement method comparison. 

We summarize similarity between the two measurement methods in the frequentist paradigm using the probability of agreement (PoA) \citep{stevens2017assessing}. The PoA used as the target of inference for our study is
\begin{equation}\label{eq:poa.target}
    \theta  = \Pr(\lvert W_{i2} - W_{i1} \rvert \le c) = \Phi\left(\dfrac{c - \gamma - (\beta-1)\mu}{\sqrt{(\beta-1)^2\sigma^2_s + \sigma^2_1 + \sigma^2_2}}\right) - \Phi\left(\dfrac{-c - \gamma - (\beta-1)\mu}{\sqrt{(\beta-1)^2\sigma^2_s + \sigma^2_1 + \sigma^2_2}}\right).
\end{equation} 
The interval $(e^{-c}, e^c)$ represents a clinically acceptable relative difference between measurements on the untransformed scale. Our study aims to precisely estimate the PoA (i.e., the probability that the two measurement methods yield clinically similar measurements) in a precision-based study.


To design this study, we define $G_r$ using the model in (\ref{eq:msdata}) with $(\gamma, \beta, \mu, \sigma_s, \sigma_1, \sigma_2) = (-0.01, 1.01,4.82,0.22, 0.06, 0.09)$ for all simulation repetitions $r$ under the conditional approach. We define the PoA using a value of $c = \log(1.2)$. This choice for $c$ implies that a relative difference of up to 20\% between the untransformed measurements is clinically acceptable. The parameters for $G_r = G$ were selected by adjusting those estimated from the \citet{bland1999measuring} data so that the true PoA is roughly $\theta = 0.9$. We consider a two-sided $100(1 - \alpha)\% = 95\%$ Wald-based confidence interval for the PoA. To ensure the bounds of the CI are between 0 and 1, the CI is first constructed on the logit scale and its endpoints are transformed (see \citet{stevens2017assessing} for details). We plan our study with the goal of obtaining an interval estimate for $\theta$ having length at most $l = 0.05$ with probability $q = 0.8$. To assess the impact of the ISU sizes, we consider $K = \{2, 3, 4, 5\}$. 

\begin{figure}[!tb]
		\includegraphics[width = \textwidth]{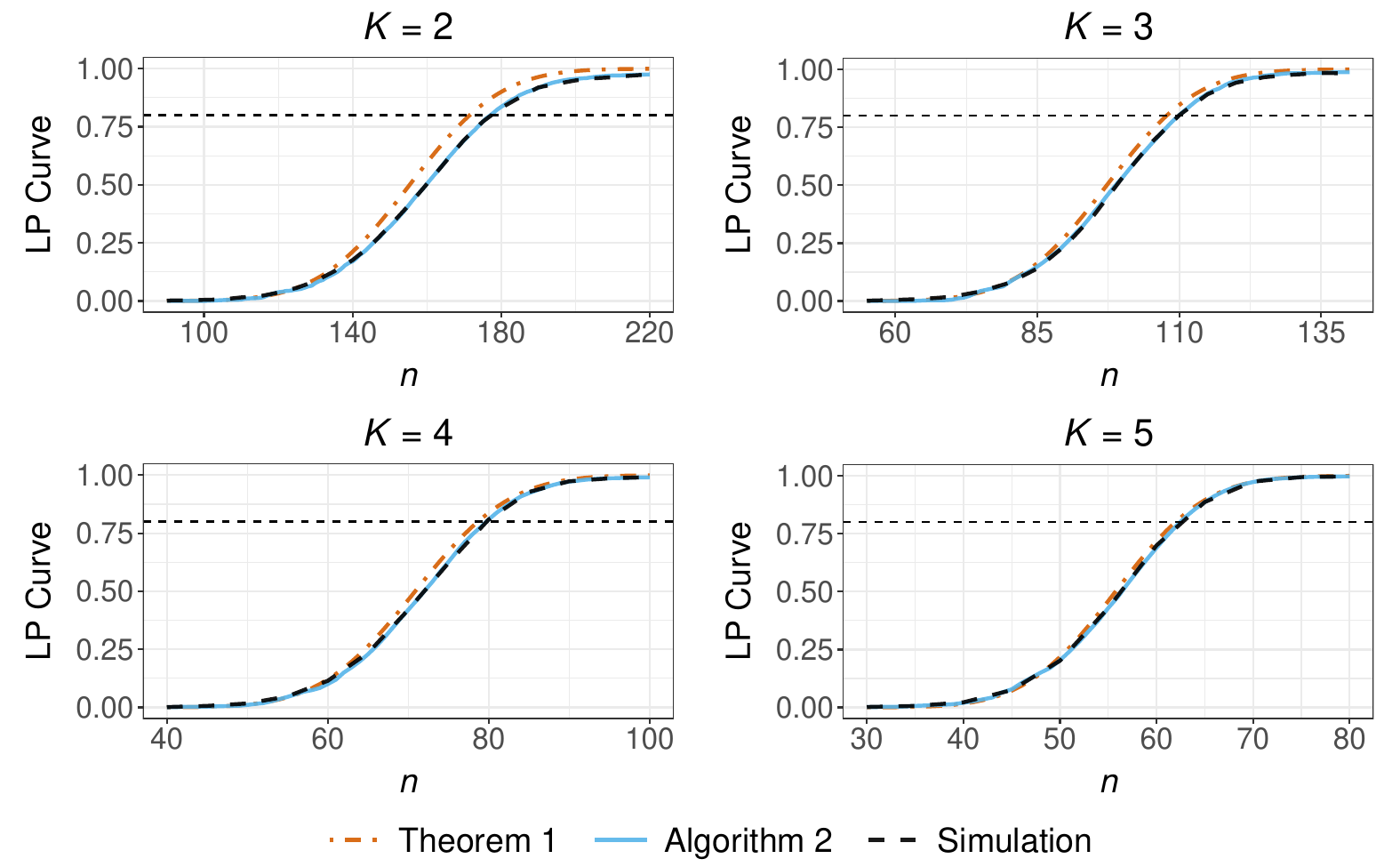} 

		\caption{\label{fig:poa} LP curves obtained using various estimation methods when the PoA is the target of inference. The horizontal dotted line represents $q = 0.8$. } 
	\end{figure}

We estimate the LP distribution using various methods for each of the scenarios described above. We first estimate the LP distribution analytically using the results from Theorem \ref{thm1}. The estimated distributions are $\mathcal{N}(155.37, 19.22)$ for $K = 2$, $\mathcal{N}(97.25, 12.49)$ for $K = 3$, $\mathcal{N}(70.86, 9.27)$ for $K = 4$, and $\mathcal{N}(55.75, 7.37)$ for $K = 5$. We then approximated the LP curves using Algorithm \ref{alg2}. Finally, we approximated the LP curve in each scenario by naively simulating the sampling distribution of CI lengths for all $n \in (\mu_l - 3\sigma_l, \mu_l + 3\sigma_l)$ that were multiples of 5, where $\mu_l$ and $\sigma_l$ are the parameters from Theorem \ref{thm1}. All sampling distributions were estimated with $m = 10^4$ simulation repetitions. Figure \ref{fig:poa} visualizes the approximated LP curves for the three estimation methods we considered. 

For all scenarios in Figure \ref{fig:poa}, the results from Algorithm \ref{alg2} align better with those obtained using naive simulation than the analytical approximation to the LP curve resulting from Theorem \ref{thm1}. Even so, the sample size recommendations based on $q = 0.8$ differ by only $n = \{6, 2, 1,1 \}$ across the three estimation methods when $K = \{2, 3, 4, 5\}$. Thus, the analytical approximation to the LP curve is reasonable for this example, which is not surprising given that the sample sizes are not too small and the CIs are relatively symmetric. Upon calculating $\Lambda_r$ and $\tau_r$ from Theorem \ref{thm1} using delta method, it took a fraction of a second to obtain the analytical approximation to each LP curve. Roughly 25 seconds on a high-performance computing server with 72 cores were required approximate the LP curve via Algorithm \ref{alg2} for each of the scenarios. It took between 2 and 6 minutes using the same computing resources to approximate the LP curves in Figure \ref{fig:poa} with naive simulation. Algorithm \ref{alg2} is more efficient because we need only estimate the sampling distribution of CI lengths at two values of $n$. To consider sample sizes in increments of 5, we required between 11 and 27 sampling distribution estimates for each curve corresponding to naive simulation. 

The sample size recommendation for $K = 2$ based on Algorithm \ref{alg2} is $n = 178$. Of course, the recommended sample size decreases as the ISU size $2K$ increases (i.e., the recommended $n$ is only 63 when $K = 5$). Practitioners can select an optimal design for this study by weighing the costs of recruiting a new participant against those associated with obtaining more repeated measurements. 


\subsection{Estimation of Incremental Net Benefit}\label{sec:study.bb}

Our next example estimates the LP distribution for a nonparametric analysis in the Bayesian paradigm. This example takes inspiration from a study on health economics that compares two interventions to treat dyspepsia \citep{willan2004incremental}. The control intervention ($j = 1$) is omeprazole, and the new intervention ($j = 2$) is omeprazole in conjunction with metronidazole and clarithromycin. Intervention $j$ has an average effectiveness $E_j$ and average cost $C_j$. The target of inference for this study is incremental net benefit, which depends on a parameter $\xi$ that denotes our willingness to pay for one unit of effectiveness. Given this parameter $\xi$, incremental net benefit is 
\begin{equation}\label{eq:inb}
\theta = (E_2 - E_1)\xi - (C_2 - C_1).
\end{equation}
In \citet{willan2004incremental}, $\xi = \$1000$ was considered, so we base sample size determination on this willingness to pay for illustration.

  We plan to collect data $\boldsymbol{w}_{ji} = (e_{ji}, c_{ji})$ for participants $i = 1, \dots, n_j$ for $j = 1, 2$. We suppose that the total sample size is $n = n_1 + n_2$ such that there is balanced randomization to the two interventions. Because the intervention assignments are randomized, $n_1$ and $n_2$ may vary slightly between simulation repetitions but always sum to $n$. Our methodology also accommodates imbalanced group sizes so long as the allocation or randomization ratio does not change with $n$. For this example, $e_{ji} \in \{0,1\}$ denotes whether participant $i$ did \emph{not} experience severe dyspeptic symptoms in the year after starting intervention $j$, and $c_{ji}$ denotes the societal cost associated with their treatment and symptoms. Under the conditional approach, we define $G_r$ such that latent variables $(x_{1i}, y_{1i})$ are simulated from a rotated CLAYTON(1.8) copula. Given these latent variables, $c_{1i}$ is the $y_{1i}$-quantile of the $\text{GAMMA}(0.7329, 0.0014)$ distribution, and $e_{1i} = 1$ if and only if $x_{1i} \le 0.37$. For intervention 2, $(x_{2i}, y_{2i})$ come from a rotated CLAYTON(0.6) copula; $c_{2i}$ is the $y_{2i}$-quantile of the $\text{GAMMA}(0.7393, 0.0016)$ distribution, and $e_{2i} = 1$ if and only if $x_{2i} \le 0.51$. This data generation process was chosen to align with summary statistics from  \citet{willan2004incremental}.

  While $\theta$ in (\ref{eq:inb}) is a function of averages, the cost distributions are quite skewed: the sickest patients have huge societal costs. In our sample, we may only have a few patients with very large costs. We therefore quantify uncertainty using the Bayesian bootstrap \citep{rubin1981bayesian} since the traditional bootstrap may leave those patients out of many resamples. The Bayesian bootstrap proceeds by drawing $\boldsymbol{v}_1 \sim \text{Dir}(\mathbf{1}_{n_1})$ and $\boldsymbol{v}_2 \sim \text{Dir}(\mathbf{1}_{n_2})$. We compute a reweighted estimate for $\theta$ as
  \begin{equation}\label{eq:bb}\theta^* = \left(\sum_{i=1}^{n_2}v_{2i}e_{2i} - \sum_{i=1}^{n_1}v_{1i}e_{1i}\right)\xi - \left(\sum_{i=1}^{n_2}v_{2i}c_{2i} - \sum_{i=1}^{n_1}v_{1i}c_{1i}\right).
  \end{equation}
  We repeat this reweighting process $M$ times. The collection of $M$  $\theta^*$ estimates creates a posterior distribution from which we can obtain a $100 \times (1 - \alpha) = 90\%$ credible interval for $\theta$ using the percentile method. We want this CI for $\theta$ to have length of at most $l = \$300$ with probability $q = 0.9$.
  
  The reweighting process above corresponds to a diffuse prior. We also consider this analysis with an informative prior. For illustration, we suppose that we have $n_0 = 150$ historical observations for the control intervention (simulated according to a process that is similar to $G_r$). We incorporate these historical observations into the analysis at a weight of half of each observation from the control intervention in the current study by simulating $\boldsymbol{v}_1 \sim \text{Dir}((\mathbf{0.5}_{n_0}, \mathbf{1}_{n_1}))$ and concatenating the historical and current data from intervention 1 in (\ref{eq:bb}). The sample size for the current study is still $n = n_1 + n_2$, and the impact of including these historical observations is negligible as $n \rightarrow \infty$. For both sets of Dirichlet priors, we consider the predictive approach for $\mathcal{G}$ such that the parameters for the rotated Clayton copulas are drawn from $\mathcal{U}[0.6, 5.4]$ and $\mathcal{U}[0.2, 1.8]$ distributions for interventions 1 and 2, respectively. Larger copula parameters imply stronger negative correlation between effectiveness and cost, resulting in greater posterior variance for $\theta$. 

\begin{figure}[!tb] \centering 
		\includegraphics[width = \textwidth]{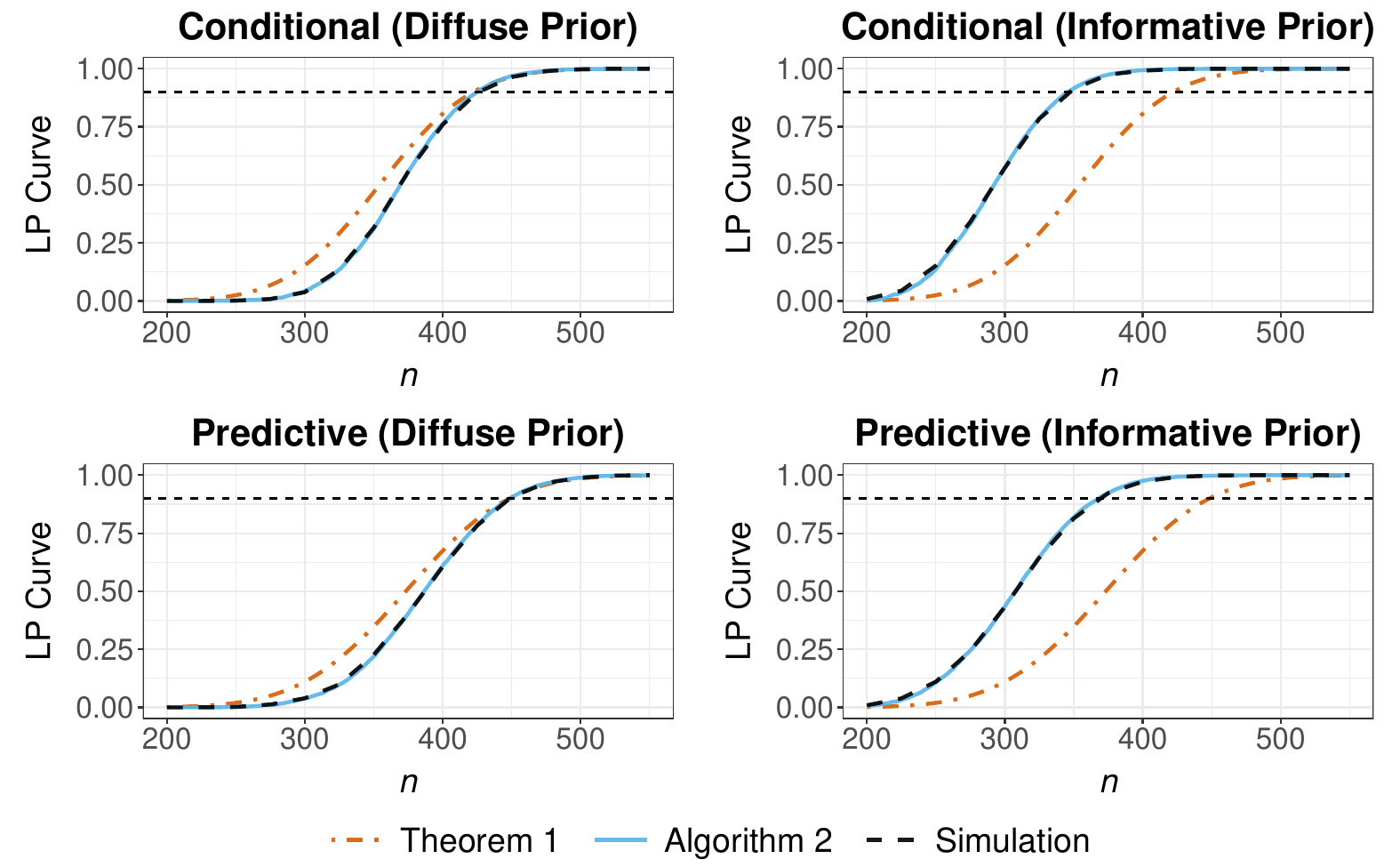} 

		\caption{\label{fig:bb} LP curves obtained using various estimation methods when incremental net benefit is the target of inference. The horizontal dotted line represents $q = 0.9$. } 
	\end{figure}

  Because $\theta$ in (\ref{eq:inb}) is an average, the result in (\ref{eq:norm1}) for the Bayesian bootstrap was proven by \citet{lo1987large}. The result in (\ref{eq:norm2}) follows by the central limit theorem since $\tau_r$ is a function of the (finite) fourth central moments of the joint distribution for effectiveness and cost across the two interventions. For each design scenario, we analytically approximated the LP curve using Theorem \ref{thm1}. This analytical approximation under the predictive approach arises from a mixture of normal distributions obtained by applying Theorem \ref{thm1} to 1000 data generation processes $G_r \sim \mathcal{G}$. We also approximated the LP curve using Algorithm \ref{alg2} with $m = 10^4$ simulation repetitions; to accommodate the predictive approach, the CI lengths were split into 10 bins based on the true $\Lambda_r$ value at each sample size before constructing the linear approximations. We lastly approximated the LP curve by simulating the sampling distribution of CI lengths at $n \in \{200, 225, \dots, 550\}$. The resulting LP curve approximations are visualized in Figure \ref{fig:bb}.

    Algorithm \ref{alg2} again yields more accurate LP curve approximation than the analytical approach. The performance of the analytical approximation resulting from Theorem \ref{thm1} is not even serviceable for the settings with the diffuse prior. Even with larger sample sizes, there is still substantial skewness in the posterior of $\theta$, resulting in a rightward shift of the true LP curve estimated using naive simulation. Because the analytical approximation to the LP curve does not account for the prior distribution, it instead overestimates the required sample size when the relatively well-specified informative prior is used ($n = 423$ vs.\ 346 for the conditional approach and $n = 449$ vs.\ 369 for the predictive approach). Overall, the LP distributions under the predictive approach exhibit greater dispersion than those under the conditional approach. That is, we require less (more) data to attain the length criterion with smaller (larger) probabilities $q$. This greater dispersion in the LP distribution is a direct result of there being greater dispersion in the sampling distribution of CI lengths under the predictive approach.

    It took one second to approximate the LP distribution under the conditional approach using Theorem \ref{thm1} with quasi-Monte Carlo integration. While considering 1000 $G_r$ under the predictive approach, this process took 15 seconds on a high-performance computing server with 72 cores. Roughly 2.5 minutes on this server were required to approximate all LP curves in Figure \ref{fig:bb} via Algorithm \ref{alg2} when using $M = 10^4$ resamples to approximate the posterior of $\theta$. On the same computing server, it took roughly 19 minutes to naively approximate the same LP curves. 

    \subsection{Estimation of Survival Distribution Quantiles}\label{sec:study.km}

    Our final example in the main text considers precision-based design for a nonparametric analysis in the frequentist paradigm. This example involves reliability analysis for the snubber component of a toaster; it is based on data from an experiment described in \citet{nelson2009accelerated}. For warranty purposes, the target of inference is the 0.25-quantile of the survival distribution, where survival is measured with respect to the number of times (cycles) that the toaster is used before the snubber component fails. In particular, $\theta = S^{-1}(0.75)$, where $S(x)$ is the survival curve evaluated at $x$ cycles. 

    We estimate the survival curve using the Kaplan-Meier estimator \citep{kaplan1958nonparametric}. We plan to construct a $100\times(1-\alpha) = 95\%$ CI for $\theta$ using the \texttt{survfit} function from the \texttt{survival} package in R \citep{therneau2026survival}. For the lower CI bound, this function returns the smallest $x$-value for which the lower bound of the pointwise 95\% CI for $S(x) \le 0.75$. For the upper CI bound, the largest $x$-value for which the upper bound of the pointwise 95\% CI for $S(x) \ge 0.75$ is returned. We construct these pointwise CIs using a log-log transformation. 

    The data for toaster $i = 1, \dots, n$ consist of $\boldsymbol{w}_i = (x_i, \delta_i)$, where $x_i$ is a time in cycles and $\delta_i \in \{0,1\}$ denotes whether this time corresponds to an observed failure time or right censoring. All toasters are considered for a maximum of 1500 cycles, and survival times are right censored if another component of the toaster fails before the snubber. For illustration, we considered two data generation processes that differ in their levels of censoring. We obtained the first data generation process $G_r$ by fitting a 10-knot spline to the Kaplan-Meier curve based on the data from \citet{nelson2009accelerated}. This spline served as a nonparametric survival curve used to generate failure times $\{y_i\}_{i=1}^n$. Censoring times $\{c_i\}_{i=1}^n$ were independently generated by condensing the $x$-axis of the 10-knot spline by a factor of 0.8. The observed event time $x_i$ and censoring indicator $\delta_i$ were based on $\min\{y_i, c_i\}$ taken after rounding to the nearest integer. This process $G_r$ reflects a scenario with a heavy censoring rate before 1500 cycles of 66\%. The second choice for $G_r$ is such that the processes to generate  $\{y_i\}_{i=1}^n$ and $\{c_i\}_{i=1}^n$ are the same, resulting in a lighter censoring rate of 50\%. Data generation in both cases is specified under the conditional approach.

The Kaplan-Meier estimator can be reconstructed as an M-estimator \citep{gu2022reconstructing}. Using this reconstruction, our choices for $G_r$ are such that the standard regularity conditions for the asymptotic normality of the M-estimator are satisfied \citep{huber2009robust}. It follows that the results in (\ref{eq:norm1}) and (\ref{eq:norm2}) hold true for the pointwise CIs for $S(x)$ that implicitly define our CI for $\theta$. As a result of this implicit relationship between the CIs for $S(x)$ and that for the target of inference, the results in (\ref{eq:norm1}) and (\ref{eq:norm2}) hold true for $\theta$ by the delta method. For this example, we want the CI for $\theta$ to have a length of at most $l = 150$ cycles with probability $q = 0.8$.

\begin{figure}[!tb] \centering 
		\includegraphics[width = \textwidth]{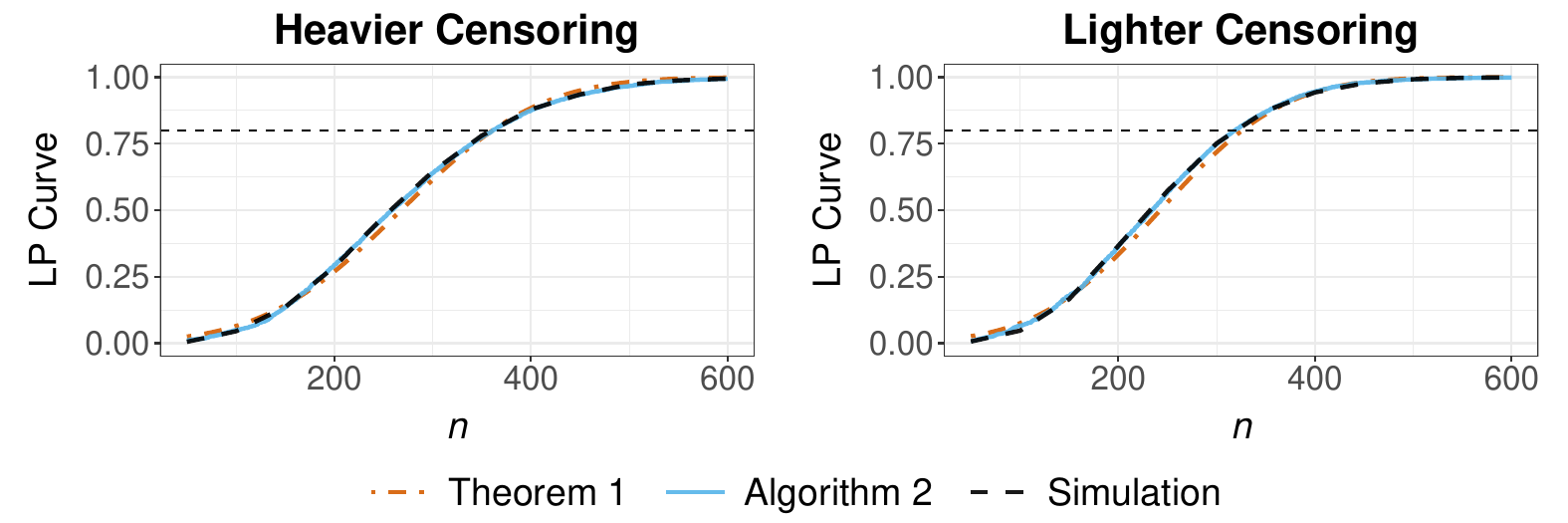} 

		\caption{\label{fig:km} LP curves obtained using various estimation methods when the 0.25-quantile of the survival curve is the target of inference. The horizontal dotted line represents $q = 0.8$. } 
	\end{figure}

   Due to the complexity of the data generation process and method for CI construction, we used a small-scale simulation with 100 repetitions at $n = 250$ to estimate the LP distribution using Theorem \ref{thm1} as described in Section \ref{sec:ssd}. We also approximated the LP curve using Algorithm \ref{alg2} with $m = 10^4$ simulation repetitions. For this example, we added $\mathcal{N}(0, 0.25^2)$ noise to the two sampling distribution estimates of CI lengths before constructing the linear approximations. This modification was required because the lower and upper bounds for all CIs -- and thus all CI widths -- are integers. Algorithm \ref{alg2} models the logarithm of the absolute difference between each CI length and the median CI length as a linear function of $\log(n)$. When a CI length is exactly the same as the median, the resulting logarithmic absolute difference of $-\infty$ cannot be used to fit a linear function. We lastly approximated the LP curve by simulating the sampling distribution of CI lengths at $n \in \{50, 100, \dots, 600\}$. The resulting LP curve approximations are visualized in Figure \ref{fig:km}.

   While Algorithm \ref{alg2} again accurately estimates the LP curve, the discrepancies in the recommended sample sizes based on Theorem \ref{thm1} are less substantial for this example ($n = 362$ vs.\ 361 for heavier censoring and $n = 319$ vs.\ 326 for lighter censoring). It took roughly half a second without parallelization to approximate the LP distribution for each scenario using Theorem \ref{thm1} and a small-scale simulation with 100 repetitions. On a high-performance computing server with 72 cores, it took 15 seconds to approximate each LP curve in Figure \ref{fig:km} via Algorithm \ref{alg2}. It took about 2 minutes to naively approximate each LP curve using the same server. 
    
    \subsection{Additional Examples}
    
    We explore the performance of our methods to approximate the LP curve with further numerical studies in Appendix B of the supplement. In Appendix B.1, these numerical studies confirm the strong performance of Algorithm \ref{alg2}, but the quality of the analytical LP curve approximation based on Theorem \ref{thm1} deteriorates when (i) CIs are based on skewed sampling or posterior distributions or (ii) informative priors are used in a Bayesian setting. We additionally consider design for two semiparametric examples based on a frequentist generalized estimating equation and a Bayesian partial likelihood in Appendices B.2 and B.3. Those examples illustrate that the results in (\ref{eq:norm1}) and (\ref{eq:norm2}) may hold true as a result of other regularity conditions that are not relevant to the examples in the main paper, underscoring the broad applicability of our methodology.  
	
	
	\section{Discussion}\label{sec:conclusion}
	
	In this paper, we developed an economical framework for study design with precision criteria. This framework determines optimal sample sizes that ensure interval estimates are narrow enough to reliably inform decision making. This framework is founded on the approximate normality of the LP distribution. We propose two methods to estimate this distribution and recommend sample sizes. One method, based on Theorem \ref{thm1}, performs well for large-sample studies and is predicated purely on asymptotic theory. The other, presented in Algorithm \ref{alg2}, flexibly and efficiently estimates the LP distribution using simulations conducted at only two sample sizes. The efficiency of this simulation-based method stems from considering a proxy for the sampling distribution of Bayesian and frequentist CI lengths based on large-sample theory. Our numerical studies illustrate that our simulation-based method to estimate the LP distribution performs well in moderate sample size settings -- even when considering informative priors or asymmetric CIs. 

  While our methodology can be broadly applied in many parametric, semiparametric, and nonparametric settings, our approach could be extended to accommodate more complex design criteria. For example, a decision-theoretic approach based on utility functions may select a sample size to balance criteria for power, precision, and financial costs \citep{calderazzo2022decision}. To facilitate economical sample size determination for such studies, our theory concerning proxy distributions for CI lengths could be combined with methods that leverage asymptotic theory to expedite simulation-based design when control over power and the type I error rate is desired \citep{hagar2025economical}.

\section*{Supplementary Material}
These materials include the proof of Theorem \ref{thm1} and additional simulation results. The code to conduct the numerical studies in the paper, along with functions to automate Algorithm \ref{alg2} that can be adapted for any design scenario, is available online: \url{https://github.com/lmhagar/LPDistribution}.

	\section*{Funding Acknowledgement}
	This work was supported by the Natural Sciences and Engineering Research Council of Canada (NSERC) by way of a PGS D scholarship, postdoctoral fellowship, and Grant RGPIN-2025-04137.
	
	
\bibliographystyle{chicago}

\end{document}